\documentclass[11pt,a4paper]{article}

\usepackage{amsthm,amsmath,amssymb,graphicx,color,bbm}

\begin{document}

\newcommand{\bc}{\begin{center}}
\newcommand{\ec}{\end{center}}
\newcommand{\be}{\begin{equation}}
\newcommand{\ee}{\end{equation}}
\newcommand{\bq}{\begin{eqnarray}}
\newcommand{\eq}{\end{eqnarray}}
\newcommand{\nn}{\nonumber}
\newcommand{\del}{\bigtriangleup}
\newcommand{\un}{\underline}
\newcommand{\Sig}{\Sigma}
\newcommand{\sig}{\sigma}
\newcommand{\til}{\tilde}
\newcommand{\sign}{\mbox{sign}}
\newcommand{\fp}[2]{\frac{\partial #1}{\partial #2}}
\newcommand{\sop}[2]{\frac{\partial^2 #1}{\partial #2^2}}
\newcommand{\thop}[2]{\frac{\partial^3 #1}{\partial #2^3}}
\newcommand{\jop}[2]{\frac{\partial^j #1}{\partial #2^j}}
\newcommand{\cp}[3]{\frac{\partial^2 #1}{\partial #2 \partial #3}}
\newcommand{\ul}{\underline}
\newcommand{\ed}{\end{document}}

\newcommand{\bea}{\begin{eqnarray}}
\newcommand{\eea}{\end{eqnarray}}
\def\kp{\kappa}
\def\m1{\mathbbm{1}}
\def\uw{\un{Z}}
\def\vw{\check{Z}}
\def\ew{e^{\beta \vw_T}_{\beta - \alpha}}
\def\eq{E^{\mathbb{Q}}_0}

\begin{titlepage}

\begin{center}
{\bf \Huge Generalizing Geometric Brownian Motion}\\
\vspace{30pt}

\begin{center}

Peter Carr, Zhibai Zhang
\vspace{20pt}

{\it Department of Finance and Risk Engineering \\
Tandon School of Engineering \\
New York University \\
12 Metro Tech Center\\Brooklyn NY 11201, USA}

\vspace{20pt}

{\tt pcarr@nyc.rr.com \\ z.zihbai@gmail.com}

\vspace{20pt}

\end{center}

\end{center}


\begin{abstract}

To convert standard Brownian motion $Z$ into a positive process,
Geometric Brownian motion (GBM) $e^{\beta Z_t}, \beta >0$ is widely used.
We generalize this positive process
by introducing an asymmetry parameter $ \alpha \geq 0$
which describes the instantaneous volatility whenever the process reaches a new low.
For our new process, $\beta$ is the instantaneous volatility
as prices become arbitrarily high.
Our generalization preserves the positivity, constant proportional drift, and tractability of GBM, while expressing the instantaneous volatility as a randomly weighted
$L^2$ mean of 
$\alpha$ and $\beta$. The running minimum and relative drawup of this process are also analytically tractable. Letting $\alpha = \beta$, our positive process reduces to  Geometric Brownian motion. By adding a jump to default to the new process, we introduce a non-negative martingale with the same tractabilities. Assuming a security's dynamics are driven by these processes in risk neutral measure, we price several derivatives including vanilla, barrier and lookback options.

\end{abstract}

\end{titlepage}

\section{Introduction}


Stochastic processes are used in option pricing models for multiple purposes.
A very common purpose is smile interpolation and extrapolation.
Given several co-terminal market quotes, the objective here is to
produce implied volatilities at a continuum of strike prices or delta levels.
A second purpose is to value path-dependent contingent claims such as
quantoed forward contracts or barrier options.
For both purposes, it is well known that
arbitrage is avoided so long as all relative price
processes
are specified as martingales under the appropriate probability measure.

In general, arbitrages can either be model-based or model-free.
An example of a model-free arbitrage is a violation of
put call parity.
An example of a model-based arbitrage is when two
European-style  futures options have different implied volatilities in
the Black model.
A martingale specification produces prices that are free of both types of arbitrage.
For example, using driftless geometric Brownian motion to describe
a futures price under the futures measure $\mathbb{Q}$
leads to both put call parity holding and
to equal implied volatilities across strikes and maturity.

Suppose that a market maker uses one martingale specification on
an initial date and then uses a different martingale specification on a second date.
For example suppose that a market maker uses
a geometric Brownian martingale with 10\% volatility
on the first date and then uses a
geometric Brownian martingale with 20\% volatility
on the second date.
The prices produced on both dates are devoid of model-free arbitrages.
For example put call parity will hold on both dates.
The prices produced on both dates do produce an arbitrage
based on the Black model being correct.
For example, if the actual volatility in the Black model is constant at 10\%, then the prices
produced on the second day allow model-based arbitrage.
If the actual volatility is instead constant at 20\%, then the prices
produced on the first day allow model-based arbitrage.
If the actual volatility is instead constant at some other value e.g. 15\%,
then the prices produced on both days allow  arbitrage
based on the Black model being correct.
However, if the Black model is not describing the risk-neutral dynamics of the
underlying, then the market maker's use
of time-inconsistent  martingale specifications  need not produce
any model-based arbitrages.
Nonetheless, the use of
time-inconsistent  martingale specifications does produce
a set of values that are devoid of
model-free arbitrages.

When the only goal is to produce values
that are devoid of
model-free arbitrages,
the only challenge to be met is to be consistent with
all of the liquid and transparent quotes.
For this purpose, time-inconsistent martingale specifications
offer greater flexibility than a time-consistent specification.
A market maker using the Black model with the same volatility on
both dates is unlikely to be able to match  the ATM quote on both dates.
In contrast,  a market maker using the Black model
with the ability to change the volatility on the second  date
is guaranteed to be able to match the ATM quotes on both dates.
In contrast this time-inconsistent  Black model does not
guarantee the ability to match
more than one option price on any given date.
When two or more simultaneous quotes differ in maturity,
and are devoid of model-free arbitrage,
one can match them by moving from the constant
volatility model to the deterministic volatility Black model.
However, when two co-terminal quotes differ in strike
and are devoid of model-free arbitrage,
one cannot necessarily match them with
the deterministic volatility Black model.
A different type of martingale specification is required to guarantee a match.

In choosing an alternative martingale specification,
it is wise to understand the reasons
behind the success of the
Geometric Brownian Martingale as the benchmark process.
Once these reasons are understood, it becomes clearer
as to which
properties of GBM should be kept
and which properties should be jettisoned.
For example, at first glance, driftless arithmetic Brownian motion
(ABM)
appears to be an attractive alternative to driftless GBM due to its
simplicity and tractability.
However, it is widely agreed that the failure
of ABM to preserve the positivity property
of GBM makes it unviable as an alternative.
It is widely argued that this positivity property of GBM
makes it a good first approximation in
describing market prices of assets whose owners enjoy
limited liability.
However, GBM has state space $(0,\infty)$ while
prices of limited liability assets occupy $[0,\infty)$.
To capture the possibility that the price of a
limited liability asset can vanish, one
can add a jump to default to a GBM, as done in \cite{m76}.

The GBM remains appropriate as a toy model
for a stock index, where it is widely agreed that zero is inaccessible.
The inaccessibility of the origin for GBM also makes it
a good toy model for an exchange rate, since if $X$ is an exchange rate,
$\frac{1}{X}$ needs to be  well defined.
For a driftless GBM, its state space and dynamics
are preserved upon inversion of the coordinate and
a change of probability measure.
In foreign exchange (FX) markets, inverting an FX rate  is a natural operation and
the change in probability measure
 corresponds to a change of numeraire.
It is highly likely that these invariance properties of GBM
explain why this stochastic process plays such a large role in the FX options
market.
If one wants to address deficiencies of GBM while retaining applicability to
FX options pricing, it stands to reason that
preserving at least some notion of invariance under inversion is crucial.
The purpose of this paper is to propose a process that generalizes GBM while
respecting invariance under inversion.
Not surprisingly, hyperbolic functions play a large role in our analysis.

It is helpful to begin by reviewing some well-known properties of GBM.
Consider an arbitrage-free market and let $\mathbb{Q}$
be an equivalent martingale measure.
Let $Z$ denote standard Brownian motion on the real line
under $\mathbb{Q}$. Consider the
process $g_t = e^{\beta Z_t}, t \geq 0$, where $\beta >0$.
Clearly, the process $g$ starts at one and stays positive forever.
From It\^o's formula:
\be
\frac{dg_t}{g_t} = \frac{\beta^2}{2}  dt + \beta  dZ_t, \qquad t \geq 0.
\label{gbm}
\ee
We say the process $g$ has constant proportional drift at
rate  $\frac{\beta^2}{2}$ and constant proportional variance at rate $\beta^2$.
The parameter $\beta$ is called the volatility.
The process $g$ is called Geometric Brownian motion.

To obtain a non-negative martingale from $g$, there are at least three approaches.
First, one can change the probability measure from $\mathbb{Q}$
to  $\tilde {\mathbb{Q}}$ by setting
$\frac{d\tilde {\mathbb{Q}}}{d \mathbb{Q}} = e^{ - \frac{\beta}{2} Z_T - \frac{\beta^2}{4}T}$.
Second, one can alternatively change the coordinate
by setting $F_t = g_t e^{- \beta^2 t/2}$.
Both of these approaches to creating a martingale
preserve the strict positivity of $g$.
If only non-negativity of the martingale is required, one can
alternatively add a  jump to default to the $g$ process with arrival rate
$\beta^2/2$.

In this paper, we propose a positive process which  generalizes
GBM $g_t = e^{\beta Z_t}, t \geq 0$ by adding
an asymmetry parameter $\alpha \geq 0$.
For our new process,
$\alpha$
describes the instantaneous volatility whenever a new low is reached.
while $\beta$ is the instantaneous volatility
whenever the process becomes arbitrarily high.
Our generalization preserves the positivity, constant proportional drift, and tractability of GBM, while expressing the instantaneous variance rate at any time as a convex combination of
$\alpha^2$ and $\beta^2$.
The model actually allows a third parameter $\gamma$ which is
the initial instantaneous volatility, and hence
is required to lie
between $\alpha$ and $\beta$.

For many options markets, three parameter models are widely used to
interpolate and extrapolate implied volatilities across strikes.
Intuitively, market participants agree that options markets
display nonzero skewness and kurtosis, but there is little discussion
about moments higher than the fourth power.
Put another way, market participants agree that it is necessary to match
some measure of level, slope, and convexity of implied volatility at the money,
but there is little discussion about the third or higher derivative  of implied volatility.

Unfortunately, our particular three parameter model is
not as flexible as some other three parameter models e.g. SABR with fixed
$\beta$ or $\rho$. As a result, our three parameter model is
only suitable for options markets where
the implied volatility slice appears to be  monotone across strike e.g. SPX or VIX.
For non-monotone slices such as when implied volatilities smile, one must alter
the model by adding e.g. stochastic volatility.
So long as the implied volatility slice  appears to be monotone across strike price,
our three parameters,  $\alpha \geq 0$,  $\beta>0$  and
$\gamma \geq 0$ have distinct and well-defined roles.
The parameter $\alpha$ controls the asymptotic implied volatility at low strikes,
while the parameter $\beta$ controls the asymptotic
implied volatility at high strikes.
The parameter $\gamma$ is used to meet an at-the-money implied volatility.

An overview of this paper is as follows.
The next section develops a new special function called
the two parameter exponential function.
The following section first uses this special function to construct a positive contibuous sub-martingale that has a constant drift.
Then we introduce a non-negative martingale
by adding a jump to default process to the sub-martingale.
This martingale has three parameters 
$\alpha \geq 0$, $\beta > 0$, and $\gamma$ between $\alpha$ and $\beta$.
This is followed by derivations of the transition PDF's for the new martingale.
The penultimate section presents closed form  valuation formulas for
contingent claims written on these martingales.
In particular, we examine vanilla options,
 lookback options and barrier options.
The final section provides both a summary of the paper
 and some suggestions for future research.

\section{Two Parameter Exponential Function}

In this section, we  construct a new
special function which we call a two parameter exponential function.
In the next section, we will use this special function to
construct our three parameter martingale.
For $\beta >0$, let $y = e^{\beta x}$ be the standard one parameter exponential function.
While the function is defined for $\beta \in \mathbb{C}$ and
$x \in \mathbb{C}$, we consider it only for
$\beta \in \mathbb{R}^{+}$ and
$x \in \mathbb{R}^+$.
The defining characteristics
of $e^{\beta x}$ are that the ratio of the function's slope to its height is constant at $\beta >0$ for all $x \geq 0$ and that the function has unit height at $x=0$ for all $\beta>0$.
Accordingly, our two parameter exponential function
will have unit height at $x=0$ for all values of its two parameters
$\alpha \geq 0$ and $\beta>0$. We will show that the
ratio of the function's slope to its height is $\alpha \geq 0$ at
$x=0$ and approaches $\beta >0$ as $x \uparrow  \infty$.
Since infinitely many functions meet just these criteria,
we further require that the ratio of the function's curvature to its height
be constant  at $\beta^2 >0$ for all $x \geq 0$.
This property also belongs to
the one parameter exponential function and
serves to uniquely\footnote{Our special function $f(x)$ solves the ordinary differential equation 
$f''(x) = \beta^2 f(x)$ on $x \geq 0$ subject to the Dirichlet boundary 
condition $f(0) = 1$ and the Neumann boundary condition $ f'(0) = \alpha$.}
  determine our two parameter exponential function.

For $x \geq 0$,  $\beta>0$, and $\alpha \geq 0$,
we define\footnote{Our function can also be expressed as $\cosh(\beta x) + \frac{\alpha}{\beta} \sinh(\beta x), x \geq 0, \alpha \geq 0, \beta>0$
and so its properties will arise as a consequence of such a representation.} the two parameter exponential function by:
\be
e^{\beta x}_{\beta - \alpha} \equiv
\frac{\beta + \alpha}{2 \beta} e^{\beta x}
+
\frac{\beta - \alpha}{2 \beta} e^{-\beta x}.
\label{gex}
\ee
Thus, the subscripted exponential is a
linear combination  of 
the ordinary exponential $e^{\beta x}$ and its reciprocal.
The $\beta - \alpha$ subscript  in
$e^{\beta x}_{\beta - \alpha} $
describes the numerator of the fraction multiplying
the reciprocal $e^{-\beta x}$.
The numerator of the fraction multiplying 
 $e^{\beta x}$ is always  the sum of the 
 asymmetry parameter $\alpha$ and the 
 scaling factor 
 $\beta$ in the ordinary exponential $e^{\beta x}$.
 The common denominator of both fractions 
 is twice this scaling factor $\beta$.
These rules uniquely expand the LHS of (\ref{gex}) into the RHS.
 
On our function's domain $x \geq 0$, the ordinary exponential 
$e^{\beta x}$
in the linear combination is larger than its reciprocal 
i.e. $e^{\beta x} \geq e^{-\beta x}$.
If $\alpha = 0$, the two fractions simplify to one half and the function is
increasing and convex. Increasing $\alpha$
increases the fraction multiplying the  larger exponential $e^{\beta x}$
and decreases the fraction multiplying the smaller exponential $e^{-\beta x}$, while keeping the value of the function at $x=0$ fixed at one.
As a result, increasing $\alpha$ causes our special function to slope up faster at every $x \geq 0$.
If $\alpha = \beta$, then the two parameter exponential
$e^{\beta x}_0$ reduces to
the one parameter exponential $e^{\beta x}$.
Thus the subscript $\beta - \alpha$
on $e^{\beta x}_{\beta - \alpha}$ is also a measure of the
deviation of our two parameter exponential function
from the one parameter exponential function.
Like the one parameter exponential function $e^{\beta x}$,
the two parameter exponential function
$e^{\beta x}_{\beta - \alpha}$
defined by (\ref{gex})  is
positive, increasing, and convex in $x$
for all $x \geq 0$ and for all
$\beta > 0$.

The derivative w.r.t. $x$ of our two parameter exponential  function is:
\be
\frac{d}{dx} e^{\beta x}_{\beta - \alpha} = \beta
e^{\beta x}_{\alpha - \beta},
\qquad \alpha \geq 0, \beta > 0, x \geq 0,
\label{d}
\ee
where:
\be
e^{\beta x}_{\alpha - \beta} \equiv
\frac{\beta + \alpha}{2 \beta} e^{\beta x}
+
\frac{\alpha - \beta}{2 \beta} e^{-\beta x},
\qquad \alpha \geq 0, \beta > 0, x \geq 0.
\label{gexm}
\ee
At $\alpha = 0$,  $e^{\beta x}_{\alpha - \beta}$ 
is the right arm of the hyperbolic sine and hence positive.
Increasing $\alpha$ increases the weight on both exponentials
and hence $e^{\beta x}_{\alpha - \beta} >0$ for all
$\alpha \geq 0, \beta > 0, x \geq 0$.
Since $\beta >0$ as well, (\ref{d}) implies that the derivative 
$\frac{d}{dx} e^{\beta x}_{\beta - \alpha}$ is positive.
Thus, the $x$-derivative
of our two parameter exponential  function
behaves the same way as the
$x$-derivative of the ordinary
exponential function w.r.t to its
scaling factor $\beta$.
Differentiating our two parameter exponential  function
w.r.t. $x$ also switches the sign on the subscript.
To convert $e^{\beta x}_{\alpha - \beta}$ on the RHS of (\ref{d})
 back into an expression involving
its cohort $e^{\beta x}_{\beta - \alpha}$, one can again
differentiate  w.r.t.  $x$.
In particular:
\be
\frac{d^2}{dx^2} e^{\beta x}_{\beta - \alpha} = \beta^2
e^{\beta x}_{\beta - \alpha},
\qquad \alpha \geq 0, \beta > 0, x \geq 0.
\label{dd}
\ee
Thus, the ratio of the function's curvature to its height is constant
at $\beta^2>0$ for all $x \geq 0$, as previously indicated.

There is an alternative way to  convert
$e^{\beta x}_{\alpha - \beta}$ back into an expression involving
its cohort $e^{\beta x}_{\beta - \alpha}$.
The appendix shows that:
\be
e^{\beta x}_{\alpha - \beta} =
\sqrt{ \left( e^{\beta x}_{\beta - \alpha} \right)^2 + \frac{\alpha^2 - \beta^2}{\beta^2}}.
\label{swap}
\ee
We now use this alternative conversion
mechanism to show that our
two parameter exponential function
sets the ratio of its slope to its height at
$\alpha$ at $x=0$.
We will also show in contrast that the ratio of its slope to its height approaches
$\beta$ as $x \uparrow \infty$.
These behaviors define the role of each parameter in our
two parameter  exponential  function.

Consider the ratio of the slope of our two parameter exponential  function
to its height:
\be
\frac{\frac{d}{dx} e^{\beta x}_{\beta - \alpha} }{e^{\beta x}_{\beta - \alpha} }
= \beta
\frac{ e^{\beta x}_{\alpha - \beta} }{e^{\beta x}_{\beta - \alpha} },
\label{ratio}
\ee
from (\ref{d}). Using (\ref{swap}) on the RHS of (\ref{ratio}),
 this ratio can also be represented as:
\be
\frac{\frac{d}{dx} e^{\beta x}_{\beta - \alpha} }{e^{\beta x}_{\beta - \alpha} }
= \beta
\frac{ \sqrt{ \left( e^{ \beta x}_{\beta - \alpha} \right)^2 + \frac{\alpha^2 - \beta^2}{\beta^2}}}{e^{\beta x}_{\beta - \alpha} }
= \beta
\sqrt{1 + \frac{\alpha^2 - \beta^2}{\beta^2 \left(e^{ \beta x}_{\beta - \alpha} \right)^2}}.
\label{null}
\ee
Bringing $\beta$ under the square root:
\be
\frac{\frac{d}{dx} e^{\beta x}_{\beta - \alpha} }{e^{\beta x}_{\beta - \alpha} }
=
\sqrt{\alpha^2  \frac{1}{ \left(e^{ \beta x}_{\beta - \alpha}\right)^2}
 + \beta^2 \left[ 1 -  \frac{1}{ \left(e^{\beta x}_{\beta - \alpha} \right)^2} \right]}.
\label{ratio1}
\ee
Since $1/\left( e^{ \beta x}_{\beta - \alpha} \right)^2 \in (0,1]$, the radicand is a convex combination
of  $\alpha^2$ and $\beta^2$.
At $x =0$, $e^{ \beta x}_{\beta - \alpha} =1$, so
$\frac{1}{ \left( e^{ \beta x}_{\beta - \alpha} \right)^2}$ also $= 1$  and the
ratio $\frac{\frac{d}{dx} e^{\beta x}_{\beta - \alpha} }{e^{\beta x}_{\beta - \alpha} } =\alpha$.
As $x \uparrow \infty$,
$e^{ \beta x}_{\beta - \alpha}  \uparrow \infty$, so
$\frac{1}{ \left(e^{ \beta x}_{\beta - \alpha} \right)^2} \downarrow 0$
 and the ratio
 $\frac{\frac{d}{dx} e^{\beta x}_{\beta - \alpha} }{e^{\beta x}_{\beta - \alpha} }$
 converges to $\beta$.

    Like the one parameter exponential function,
our two parameter exponential function has an explicit inverse.
To derive it, let:
\be
y =  e^{\beta x}_{\beta- \alpha} =
\frac{\beta + \alpha}{2 \beta} e^{\beta x}
+
\frac{\beta - \alpha}{2 \beta} e^{-\beta x},
\qquad x \geq 0, \alpha \geq 0, \beta>0.
\label{y}
\ee
We need to solve for $x$ as a function of $y$.
Multiplying (\ref{y}) by $\beta e^{\beta x}$ leads to a quadratic
function of $e^{\beta x}$:
\be
\frac{\beta + \alpha}{2} e^{2 \beta x}
- \beta y e^{\beta x}
+ \frac{\beta - \alpha}{2}
= 0,
\qquad x \geq 0, \alpha \geq 0, \beta>0.
\label{q}
\ee
By the quadratic root formula:
\be
e^{\beta x}
=
\frac{
\beta y + \sqrt{ \beta^2 y^2 - (\beta^2 - \alpha^2)}
}{\beta + \alpha},
\qquad x \geq 0, \alpha \geq 0, \beta>0,
\label{qrf}
\ee
where we have chosen $+$ in $\pm$ since $e^{\beta x}>0$.
Solving for $x$:
\be
x =
\frac{1}{\beta} \ln
\frac{
\beta y + \sqrt{\alpha^2 +  \beta^2 (y^2 - 1)}
}{\beta + \alpha},
\qquad x \geq 0, \alpha \geq 0, \beta>0.
\label{inv0}
\ee
Hence, for $y \geq 1$, the function on the RHS of
(\ref{inv0}) is the explicit inverse of our two parameter exponential function.

Notice that from (\ref{qrf}):
\be
\frac{\beta + \alpha}{2 \beta} e^{\beta x}
= \frac{y}{2} + \sqrt{ \frac{y^2}{4}  - \frac{\beta^2 - \alpha^2} {4 \beta^2} },
\label{qrf1}
\ee
where we observe from (\ref{y})   that  $\frac{\beta^2 - \alpha^2} {4 \beta^2}$ is
just the product of the
two terms which sum to $y$.
Equation (\ref{qrf1}) is an explicit formula that maps $y$ to the first term in the sum
(\ref{y}) defining it.
When $x=0$ and $\alpha = 0$, this first term has the same size
of $\frac{1}{2}$  as the second term, but otherwise, the first term is larger.
To obtain an
explicit formula that maps $y$ to the smaller term in the sum defining it,
notice that multiplying (\ref{y}) by $\beta e^{-\beta x}$ leads to a quadratic
function of $e^{-\beta x}$:
\be
\frac{\beta - \alpha}{2} e^{-2 \beta x}
- \beta y e^{\beta x}
+ \frac{\beta + \alpha}{2}
= 0.
\label{q3}
\ee
By the quadratic root formula:
\be
e^{- \beta x}
=
\frac{
\beta y - \sqrt{ \beta^2 y^2 - (\beta^2 - \alpha^2)}
}{\beta - \alpha},
\label{qrf3}
\ee
where now we have chosen $-$ in $\pm$ since $e^{-\beta x} <1$.
Hence:
\be
\frac{\beta - \alpha}{2 \beta} e^{- \beta x}
=
\frac{y}{2} - \sqrt{ \frac{y^2}{4}   - \frac{\beta^2 - \alpha^2} {4 \beta^2} }.
\label{qrf4}
\ee
This equation is an explicit formula that maps $y$ to the
last smaller term in the sum (\ref{y}) defining it.

For the one parameter exponential function $y = e^{\beta x}, x \geq 0, \beta > 0$,
adding one to the input variable $x$ causes the  output variable $y$
to grow by the factor $e^{\beta}>1$. We say the exponential function turns addition into multiplication.
For our two parameter exponential function defined by (\ref{y}),
adding one to the input variable $x$ causes the  output variable $y$
to grow as follows. First,
split $y$ into its larger term involving $e^{\beta x}$
given explicitly by (\ref{qrf1})
and  its smaller term involving
$e^{-\beta x}$,
given explicitly by (\ref{qrf4}).
Next, grow the larger term by a factor
$e^{\beta} >1$ and shrink the smaller term by
a factor  $e^{-\beta} \in (0,1)$.
Finally, add the two altered terms together
to obtain the new value of $y$.
We say the two parameter exponential function turns addition into
a blend of multiplication and division.

\section{ Constructing a 3 Parameter Non-Negative Continuous Martingale}
In this section, we  use
the two parameter exponential function
constructed in the last section
to define a new three parameter non-negative continuous martingale
denoted by $F_t$.
Recall that
to create a driftless GBM $F^b$ ,
one first creates an auxiliary positive continuous process
$g_t = e^{\beta Z_t}$ with constant positive drift of $\beta^2/2$
and then one corrects for this constant drift by setting
$\frac{F^b_t}{F^b_0} = g_t e^{-\beta^2 t/2}$.
We will mimic this construction in the next subsection by
first constructing an auxiliary positive continuous process $G$ with
positive constant drift of $\beta^2/2$.
The following subsection then
corrects for this constant drift by 
adding a jump to default process.

\subsection{ Constructing a Positive Continuous Process with Constant Drift}
Let $0$ be the valuation time and let $Z$ be
a standard Brownian motion $Z$
under $\mathbb{Q}$ whose value at $t=0$ is  $Z_0 = 0$ as usual.
We allow
$Z$ to exist prior to time 0.
Let $t_0 \leq 0$ and we suppose that
$Z$ exists for all $t \geq t_0$.
For $t \geq t_0$,
let $\underline{Z}_t \equiv \stackrel{\inf}{\scriptstyle s \in [t_0,t]} Z_s$ denote the running minimum of the standard Brownian motion $Z$ under $\mathbb{Q}$.
Notice that $Z$'s path  monitoring begins at time $t_0 \leq 0$, so
$\underline{Z}_0 \leq 0$.
For $t \geq t_0$,
let $\check{Z}_t \equiv Z_t - \underline{Z}_t$ denote $Z$'s running drawup process. Let:
\be
\check{G}_t = e^{\beta \check{Z}_t}_{\beta - \alpha},
\qquad t \geq t_0, \beta >0,
\label{gdu1}
\ee
be a new stochastic process with state space $[1,\infty)$.

Recall  that setting $\alpha $ to zero reduces the two parameter exponential 
$e^{\beta x}_{\beta - \alpha}, x \geq 0, \beta >0$
to the ordinary exponential $e^{\beta x}, x \geq 0, \beta >0$. 
The GBM $e^{\beta Z_t}$  and the processes $\cosh(\beta Z_t)$, $\cosh(\beta |Z_t|)$,
and  $\cosh(\beta \check{Z}_t)$
all grow in expectation at the rate  $\beta^2/2$.
The hyperbolic cosine is a simple average of the increasing exponential 
$e^{\beta x}, x \geq 0, \beta >0$, 
and its reciprocal.   
When the asymmetry parameter $\alpha$ is made positive,
this simple average is replaced with an asymmetric average putting more weight on 
the increasing exponential. The effect on the mean of this skewing is 
the same as the effect on the mean of the GBM $e^{\beta Z_t}$
 if $Z$ behaved asymmetrically just when visiting its minimum $\bar{Z}$.
 In particular, if $Z$ is interpreted as a the limit of a scaled random walk, 
 then putting greater probability on rising above the minimum raises the mean 
 growth rate of $e^{\beta Z_t}$  above $\beta^2/2$. 
 Let $\hat{Z}$ denote  this skewed Brownian motion. 
 The effect on the mean of 
 $e^{\beta \hat{Z}}_t$ of 
 this rarely imposed asymmetry can be removed by multiplying
  by $e^{\alpha \bar{Z}_t}$.
 We will similarly remove the effect on the mean of 
 $\check{G}_t $ of replacing 
 $\cosh(\beta \check{Z}_t)$
 with
$e^{\beta \check{Z}_t}_{\beta - \alpha},  \beta >0, \alpha \geq 0$
 by multiplying $\check{G}_t $
  by $e^{\alpha \bar{Z}_t}$.

We introduce a new parameter $\gamma$ which will be used to
determine the value of
$\check{G}_t$ at $t=0$. We require that $\gamma$ be between $\alpha$ and $\beta$.
For technical reasons, we allow   $\gamma = \alpha$, but we do not allow
$\gamma = \beta$. This allows us to set:
\be
\check{G}_0 = \sqrt{\frac{\alpha^2 - \beta^2}{\gamma^2 - \beta^2} }.
\label{id}
\ee
The radicand  is $\geq 1$ and hence so is $\check{G}_0$.
We next use (\ref{inv0}) to set $\check{Z}_0$:
\be
\check{Z}_0 =
\frac{1}{\beta} \ln
\frac{
\beta \check{G}_0 + \sqrt{\alpha^2 +  \beta^2 [\check{G}_0^2 - 1]}
}{\beta + \alpha},
\qquad x \geq 0, \alpha \geq 0, \beta>0.
\label{inv1}
\ee
Since $\check{G}_0 \geq 1$,  $\check{Z}_0 \geq 0$.
At each $t \geq 0$, $\check{G}_t \geq 1$ defined in (\ref{gdu1})
is increasing in its driver $\check{Z}_t \geq 0$.
Equation (\ref{inv}) implies that (\ref{gdu1}) can be explicitly inverted:
\be
\check{Z}_t
=
\frac{1}{\beta} \ln \left(
\frac{
\beta \check{G}_t + \sqrt{ \alpha^2 + \beta^2 [\check{G}_t^2-1]}
}{\alpha + \beta} \right), \qquad t \geq t_0, \alpha \geq 0, \beta > 0.
\label{inv}
\ee

We next set $\underline{Z}_0 = - \check{Z}_0$ so that
$Z_0 \equiv  \underline{Z}_0 + \check{Z}_0 = 0$.
With $\underline{Z}_0$ determined at some non-positive value, let:
\be
\underline{G}_t =
e^{ \alpha \underline{Z}_t}
\qquad t \geq t_0, \alpha \geq 0,
\label{Gmindef}
\ee
be a super-martingale with state space $(0,1]$.
The process $\underline{G}_t \in (0,1]$ defined in (\ref{Gmindef})
is increasing in its driver $\underline{Z}_t \leq 0$,
For $\alpha > 0$, (\ref{Gmin}) can be explicitly inverted:
\be
\underline{Z}_t =  \frac{1}{\alpha} \ln \underline{G}_t,\qquad t \geq t_0.
\label{Wmin}
\ee

For $\alpha \geq 0, \beta > 0$ and $\gamma$ between them, let:
\be
G_t = \underline{G}_t \check{G}_t,
\qquad t \geq 0,
\label{Gdef}
\ee
be our auxiliary continuous process with state space $(0,\infty)$.
We claim that $\underline{G}_t = \stackrel{\inf}{\scriptstyle s \in [t_0,t]} G_s$.
In words, we claim that the super-martingale $\underline{G}_t \in (0,1]$ defined in (\ref{Gmindef})
is just the running minimum of the $G$ process defined in (\ref{Gdef}).
To see why, note that substituting (\ref{gdu1}) and (\ref{Gmindef}) in (\ref{Gdef})
implies that for $\alpha \geq 0, \beta > 0$ and $\gamma$ between them:
\be
G_t =
e^{ \alpha \underline{Z}_t}
e^{\beta \check{Z}_t}_{\beta - \alpha},
\qquad t \geq t_0.
\label{Gdef1}
\ee
Since $\underline{Z}$ only declines when $\check{Z}=0$:
\be
\stackrel{\inf}{\scriptstyle s \in [t_0,t]} G_s =   e^{ \alpha \underline{Z}_t},
\qquad t \geq t_0,\alpha \geq 0, \beta > 0,
\label{Gmin}
\ee
which matches the defining equation (\ref{Gmindef}) for $\underline{G}_t$.
Hence $\underline{G}_t $ is the running minimum of the $G$ process
defined in (\ref{Gdef}).
Since $\underline{G}_t$ has state space $(0,1]$,  $G$ is positive forever.
From (\ref{Gdef}):
\be
\check{G}_t = \frac{G_t}{\underline{G}_t}, \qquad t \geq 0,
\label{gdu}
\ee
so $\check{G}$ is the relative drawup process of $G$.

Applying It\^{o}'s formula to (\ref{gdu1}), (\ref{d}) implies that:
\be
d\check{G}_t = \beta  e^{\beta \check{Z}_t}_{\alpha - \beta}
d\check{Z}_t + \frac{\beta^2}{2} e^{\beta \check{Z}_t}_{\beta - \alpha} d \langle \check{Z} \rangle_t,
\qquad t \geq t_0.
\label{gdu1a}
\ee
Thus the increments of $\check{G}_t$ depend on the increments of
$\check{Z}_t$ and the squared increments of  $\check{Z}_t$.
Since $\underline{Z}$ is a process of bounded variation, it has zero quadratic variation and hence:
\be
\langle \check{Z} \rangle_t  =
\langle Z - \underline{Z} \rangle_t =  \langle Z \rangle_t   = t,
\qquad t \geq t_0.
\label{qv}
\ee
Substituting (\ref{swap}) and (\ref{qv}) in (\ref{gdu1a}) implies
that  the coefficients just depend on
$e^{\beta \check{Z}_t}_{\beta - \alpha}$:
\be
d\check{G}_t =
 \frac{\beta^2}{2}
e^{\beta \check{Z}_t}_{\beta - \alpha} dt
+ \beta
\sqrt{ \left( e^{\beta \check{Z}_t}_{\beta - \alpha} \right)^2 + \frac{\alpha^2 - \beta^2}{\beta^2}}
d\check{Z}_t,
\qquad t \geq t_0.
\label{gdu1b}
\ee
Substituting (\ref{gdu1}) in (\ref{gdu1b}) implies
that $\check{G}$ solves
the following
stochastic differential equation
(SDE):
\be
d\check{G}_t =
 \frac{\beta^2}{2}
\check{G}_t
 dt
+
\sqrt{\alpha^2 +  \beta^2 \left[ \left(\check{G}_t\right)^2   - 1 \right]}
d\check{Z}_t,
\qquad t \geq t_0.
\label{gdu1c}
\ee
This SDE is univariate since the
coefficients for $\check{G}_t$ just depend on $\check{G}_t$.
Dividing by $\check{G}_t$ implies:
\be
\frac{d\check{G}_t}{\check{G}_t} =
 \frac{\beta^2}{2} dt
+
\sqrt{ \alpha^2  \frac{1}{\check{G}^2_t}  +
  \beta^2 \left[1 -   \frac{1}{\check{G}^2_t}    \right]}
d\check{Z}_t,
\qquad t \geq t_0.
\label{gdu1d}
\ee
Hence, $\check{G}$ solves the above simple SDE when the two drivers
are $t$ and $\check{Z}$. To determine the coefficients of $Z_t$ and $\un{Z}$, note that
substituting $d\check{Z}_t = dZ_t - d \underline{Z}_t$ in (\ref{gdu1d}) implies:
 \be
\frac{d\check{G}_t}{\check{G}_t}
=    \frac{\beta^2}{2} dt
+
\sqrt{ \alpha^2  \left( \frac{1}{\check{G}_t} \right)^2 +
  \beta^2 \left[1 -  \frac{1}{\check{G}^2_t}    \right]}
\left(dZ_t - d \underline{Z}_t\right),
\qquad t \geq t_0.
\label{it1g}
\ee
Since $\underline{Z}$ only decreases when $\check{G} = 1$, the
net coefficient of $d\underline{Z}$ in (\ref{it1g}) is zero.
As a result,
$\check{G}$ also solves the following
SDE:
\be
\frac{d\check{G}_t}{\check{G}_t} =
- \alpha d \underline{Z}_t
+ \frac{\beta^2}{2} dt
+ \sqrt{ \alpha^2  \frac{1}{\check{G}^2_t}
+ \beta^2 \left[ 1 -  \frac{1}{\check{G}^2_t} \right] }dZ_t,
\qquad t \geq t_0.
\label{Gvsde}
\ee
The coefficient of $dZ_t$ in 
(\ref{Gvsde}) is the instantaneous lognormal volatility of $\check{G}$,
which  is a randomly weighted $L^2$ mean of $\alpha$ and $\beta$.
This form is clearly just a consequence of (\ref{ratio1}).
Since $\left( \frac{1}{\check{G}_t} \right)^2 \in (0,1]$, the
instantaneous lognormal variance rate of
$\check{G}_t$ is just a convex combination of
$\alpha^2$ and $\beta^2$.
When $Z$ is at its minimum $\underline{Z}$,
$\check{Z} = 0$,
and hence $\check{G}=1$.
At such times, (\ref{Gvsde}) implies that the
instantaneous volatility of $\check{G}$ is $\alpha$.
In contrast, as the difference between $Z$
and its minimum $\underline{Z}$ approaches infinity,
$\check{G}$ also approaches infinity, and (\ref{Gvsde}) implies that
the instantaneous volatility of $\check{G}$ approaches $\beta$.
These results clearly follow from the
behavior of our two parameter exponential function
$e^{\beta x}_{\beta - \alpha}$
at $x=0$ and at $x=\infty$.

 
The dynamics in (\ref{Gvsde}) clearly depend on our 
first two parameters 
$\alpha$ and $\beta$, which are 
the respective instantaneous volatilities of 
$\check G$ at $\check G$'s extremes of one and infinity.
To interpret our third parameter $\gamma$, note that
squaring both sides of (\ref{id}) implies that:
\be
\check{G}_0^2 = \frac{\alpha^2 - \beta^2}{\gamma^2 - \beta^2}.
\label{id2}
\ee
Cross multiplying and re-arranging:
\be
\gamma^2 \check{G}_0^2 = \alpha^2 - \beta^2 + \beta^2 (\check{G}_0)^2.
\label{id3}
\ee
Dividing by $\check{G}_0^2$ and taking the square root implies:
 \be
\gamma = \sqrt{ \alpha^2   \frac{1}{\check{G}^2_0}  +
  \beta^2 \left[1 -  \frac{1}{\check{G}^2_0}    \right]}.
\label{invol}
\ee
Comparing  (\ref{invol}) to the volatility in (\ref{Gvsde}) evaluated at $t=0$ implies that
our third parameter $\gamma$ is just the initial volatility of $\check{G}$.

We next determine the dynamics of the $G$ process, which
(\ref{Gdef}) defined as the product:
 \be
G_t =  \underline{G}_t \check{G}_t
\qquad t \geq t_0,
\label{gdut}
\ee
for $\alpha \geq 0, \beta > 0$ and $\gamma$ between them.
It\^{o}'s formula implies that:
 \be
\frac{dG_t}{G_t}
=  \frac{d\underline{G}_t}{\underline{G}_t}
+ \frac{d\check{G}_t}{\check{G}_t}
=  \alpha d \underline{Z}_t
+ \frac{d\check{G}_t}{\check{G}_t},
\qquad t \geq t_0,
\label{it}
\ee
since $\underline{G}_t = e^{ \alpha \underline{Z}_t}$.
Substituting in (\ref{gdu1d}) implies that
$G$ solves the following
SDE:
\be
\frac{dG_t}{G_t} = \frac{\beta^2}{2} dt
+ \sqrt{ \alpha^2 \left( \frac{\underline{G}_t}{G_t} \right)^2
+ \beta^2 \left[ 1 - \left(\frac{\underline{G}_t}{G_t}  \right)^2 \right] }dZ_t,
\qquad t \geq t_0,
\label{Ssde}
\ee
since $\frac{1}{\check{G}_t} = \frac{\underline{G}_t}{G_t}$.

Like the $\check{G}$ process,
the $G$ process has constant proportional drift at rate
$\frac{\beta^2}{2}$.
Unlike the SDE (\ref{Gvsde}) for $\check{G}$, the SDE
(\ref{Ssde})  for
$G$ has coefficients that depend on the auxiliary process
$\underline{G}$.
Since $\left( \frac{\underline{G}_t}{G_t} \right)^2 \in (0,1]$,
the lognormal variance rate of $G$
is also a convex combination of $\alpha^2$ and $\beta^2$.
When $G_t =  \underline{G}_t$, the
$G$ process behaves locally like a
GBM with constant proportional drift rate
 $\frac{\beta^2}{2}$
and constant proportional variance rate $\alpha^2$.
As $G$ rises above $\underline{G}_t$,
the lognormal variance rate moves towards
$\beta^2$ and asymptotes to this value in the limit as
$G \uparrow \infty$.

Substituting $\frac{1}{\check{G}_0} = \frac{\underline{G}_0}{G_0}$
in (\ref{invol}) implies that:
 \be
\gamma = \sqrt{ \alpha^2  \left( \frac{\underline{G}_0}{G_0} \right)^2 +
  \beta^2 \left[1 -  \left( \frac{\underline{G}_0}{G_0} \right)^2   \right]}.
\label{invol1}
\ee
Evaluating the coefficient of $dZ_t$ in (\ref{Ssde})
at $t=0$, (\ref{invol1}) implies that
the instantaneous lognormal volatility of $G$ is $\gamma$.
Hence, the three parameters
$\alpha, \gamma$, and $\beta$
can be respectively interpreted as the instantaneous
lognormal volatility of $G$ at each new low,
at the initial time, and at infinitely high values of $G$.

The bivariate transition PDF of the pair $(\underline{Z},\check{Z})$ is known in
closed form and is given in \cite{c14}.
Since $\underline{G}$ and $\check{G}$ are each just univariate, increasing,
explicitly invertible transformations of
$\underline{Z}$ and $\check{Z}$ respectively,
it follows that
the bivariate transition PDF of the pair $(\underline{G},\check{G})$
can easily be obtained in closed form.

Recall  from (\ref{Gdef1}) that:
\be
G_t =
e^{ \alpha \underline{Z}_t}
e^{\beta \check{Z}_t}_{\beta - \alpha},
\qquad t \geq t_0.
\label{Gdef1r}
\ee
As $\beta \downarrow 0$, the
$G$ process becomes driftless and
two parameter exponential
function $e^{\beta x}_{\beta - \alpha}$ in (\ref{Gdef1r})
converges  to the linear function
$1 + \alpha x$.
As a result, the process $G$ converges to the martingale
$F$ in \cite{c14} in this limit when $F_0 =1$.
Setting $\alpha = \beta$ in (\ref{Gdef1r}), the two parameter
exponential reduces to the one parameter exponential
and hence:
\be
G_t =
e^{ \beta \underline{Z}_t}
e^{\beta \check{Z}_t}
= e^{\beta (\underline{Z}_t + \check{Z}_t)}
= e ^{\beta Z_t},
\qquad t \geq t_0.
\label{Gdef1r1}
\ee
Thus, the $G$ process generalizes the exponential of
standard Brownian motion, by adding parameters $\alpha$ and
$\gamma$. 

Being a sub-martingale, the $G$ process can be used directly to model spot price (e.g. spot FX rates) and price derivatives written on $G$ in risk neutral measure. For this purpose, we introduce a new sub-martingale process 
\bea
F_t=F_0G_t, \quad t\geq t_0,\label{submgl}
\eea
where $F_0>0$ is the initial value of the process. Like $G$, $F$ is positive and has a positive drift. Note that the positivity of the drift of $G$ is not a binding restriction due to the international put-call equivalence \cite{g83}. For instance, if a positive process $S_t$ has a negative drift, one can use it to model the inverse of a process that has a positive drift via $F_t=\frac 1{S_t}$. For derivatives on future price, the underlying security is required to be driven by a martingale in the risk neutral measure for derivative pricing. In the next subsection we introduce a new martingale process from $G$ by adding a jump to default process which has a negative drift. However, one should interpret the sub-martingale Eqn (\ref{submgl}) and the new martingale as dynamics of two different securities, instead of spot and future prices of one security.

\subsection{Constructing a Non-Negative Martingale via Jump to Default}

For $\alpha \geq 0, \beta >0$, and for $\gamma$ between them,
 the $G$ process constructed in the last subsection
 starts at one and has constant positive  drift $\frac{\beta^2}{2}$.
In this section, we change the starting point to $F_0 >0$ and interpret this
positive drift as compensation for a possible jump to zero.
This allows us to
construct a tractable non-negative martingale
$F$ which starts at $F_0$. Let $N_t$ be a standard Poisson process with
arrival rate $\frac{\beta^2}{2}$ under $\mathbb{Q}$.
For $F_0 >0$, let:
 \be
F_t = F_0 G_t \m1_{N_t = 0},
\qquad t \geq t_0.
\label{fjtd}
\ee
be a non-negative process started at $F_0>0$.
Then $F$ is a $\mathbb{Q}$ martingale which drifts up at
the constant rate
$\frac{\beta^2}{2}$ in order to compensate for
a possible jump to zero. Once $F$ hits zero, it is absorbed there.
Let:
\be
\un{F}_t = \stackrel{\inf}{\scriptstyle s \in [t_0,t]} F_s,
\qquad t \geq t_0
\label{Fbar}
\ee
be the running minimum of $F$. Let $\tau$ be the exponentially distributed
random time at which $F$ jumps to zero.
For $t \in [t_0, \tau)$, (\ref{fjtd}) implies:
\be
\underline{F}_t = F_0 \un{G}_t.
\label{unfjtd}
\ee
Dividing (\ref{unfjtd}) by (\ref{Fbar}) implies that for $t \in [t_0, \tau)$:
\be
\frac{\underline{F}_t}{F_t}=
\frac{\underline{G}_t}{G_t}.
\label{rf}
\ee
 As a result, the SDE for $F$ is:
\be
dF_t  =
F_{t-} \left[ \sqrt{ \alpha^2 \left( \frac{\underline{F}_{t-}}{F_{t-}} \right)^2
+ \beta^2 \left[ 1 - \left(\frac{\underline{F}_{t-}}{F_{t-}}  \right)^2 \right] }dZ_t
- \left(d N_t -  \frac{\beta^2}{2} dt \right) \right],
\qquad t \geq t_0.
\label{Fsdejtd}
\ee

Substituting (\ref{Gdef})  in (\ref{fjtd})
implies that $F_t$ can be related to the
contemporaneous values of the pair
$(\underline{Z},\check{Z})$ and $N_t$:
 \be
F_t
= F_0
 e^{ \alpha \underline{Z}_t }
e^{\beta \check{Z}_t}_{\beta - \alpha}
\m1_{N_t = 0},
\qquad t \geq t_0.
\label{new1f}
\ee
The price relative $\frac{F_t}{F_0}$ is a non-negative
martingale started at one. From (\ref{new1f}), this price relative
decomposes into the product of a positive strict supermartingale started at one,
$e^{ \alpha \underline{Z}_t } \m1_{N_t = 0}$ and a
positive strict submartingale started at one, namely
$\check{G}_t =
e^{\beta \check{Z}_t}_{\beta - \alpha}$.

If $\alpha = \beta$, then the two parameter
exponential function
$e^{\beta x}_{\beta - \alpha}$
in (\ref{new1f})
reduces to the one parameter exponential function
$e^{\beta x}$,
and hence (\ref{new1f}) simplifies to:
\be
F_t
= F_0
e^{ \beta \underline{Z}_t}
e^{\beta \check{Z}_t}\m1_{N_t = 0}
= F_0
e^{ \beta (\underline{Z}_t +  \check{Z}_t)}\m1_{N_t = 0}
=
e^{ \beta Z_t}\m1_{N_t = 0},
\qquad t \geq t_0,
\label{new2f}
\ee
which is  GBM with jump to default.
When $\beta \rightarrow 0$, then (\ref{new1f}) asymptotes to:
 \be
F_t
\rightarrow F_0
 e^{ \alpha \underline{Z}_t}  (1 + \alpha  \check{Z}_t),
\qquad t \geq t_0,
\label{new3f}
\ee
which is a two parameter positive continuous martingale.
Setting $\gamma = \alpha$ further reduces
$F$ to the one parameter
positive continuous martingale in \cite{c14}. 

From \cite{c14}, the bivariate transition PDF of
the Brownian Minimum and Brownian Drawup:
\bea
\mathbb{Q}_t  \{\un{Z}_T \in dj,\check{Z}_T \in d\check{k}; \un{Z}_t = \un{Z}, \check{Z}_t = \check{Z} \} = b(j,\check{k};w,T-t) dj d\check{k} \nn\\
 b(j,\check{k};w,T-t) \equiv \sqrt{ \frac{2}{\pi (T - t)^3}} (\check{k}-j + w)e^{- \frac{(\check{k}-j+w)^2}{2(T-t)}},\qquad j < \ul{w}, \quad \check{k} \geq 0, \label{jden}
\eea
 where $w = \un{Z} + \check{Z}$ and $\ul{w}=\ul{Z}$. Note that in a special case when $\ul{Z}_T=\ul{Z}_t$, the bivariate transition PDF becomes a univariate one:
\bea 
\tilde{\mathbb{Q}}_t  \{\un{Z}_T = \ul{Z}_t,\check{Z}_T \in d\check{k}; \un{Z}_t = \un{Z}, \check{Z}_t = \check{Z} \} = \tilde{b}(\check{k};w,T-t) d\check{k} \nn\\
 \tilde{b}(\check{k};w,T-t)\equiv \sqrt{ \frac{2}{\pi (T - t)}} \left( e^{-\frac{\check{k}^2}{2(T-t)}} - e^{-\frac{(\check{k}+w-\ul{w})^2}{2(T-t)}} \right), \qquad \check{k} \geq 0\,.\label{jden1}
\eea

Next we construct the bivariate transition PDF for the double-exponential process (\ref{new1f}).  Let $\un{F}^s_T$  be the minimum of $F$ at $T$ conditional on surviving to $T$.
Similarly, let  $\check{F}^s_T$
be the drawup of $F$ at $T$, conditional on surviving to $T$.
The bivariate transition PDF of
the Brownian Minimum and Brownian Drawup can be used to
derive the bivariate PDF of the pair $(\un{F}^s_T, \check{F}^s_T)$,
conditional both on surviving to $T$ and on $(\un{F}^s_t,\check{F}_t) = (\un{F}, \check{F})$.
 For $J \in (0,F_0]$, and $\check{K} \geq 1$, we seek:
$$\mathbb{Q}  \{\un{F}^s_T \in dJ, \check{F}^s_T \in d\check{K} |N_T = 0, \un{F}^s_t=\un{F}, \check{F}_t = \check{F} \}.$$
In other words, we wish to know the bivariate conditional PDF when we change variables from $(j,\check{k})$ to:
$$ (J,\check{K}) = (F_0  e^{\alpha j}, e^{\beta \check{k}}_{\beta - \alpha}).$$
Let $j(J)$ be the inverse of $J = F_0  e^{\alpha j}$:
\be
j(J) = \frac{1}{\alpha} \ln \left( \frac{J}{F_0} \right), \qquad J \in (0,F_0].
\label{jdef0}
\ee
Similarly, let $\check{k}(\check{K})$ be the inverse of $\check{K} = e^{\beta \check{k}}_{\beta - \alpha}$:
\be
\check{k}(\check{K}) = \frac{1}{\beta} \ln \left[ \frac{\beta \check{K} + \sqrt{\alpha^2 + \beta^2(\check{K}^2-1)} }{\alpha + \beta} \right], \qquad \check{K} \geq 1.
\label{jdef}
\ee
The determinant of the Jacobian for this change of variables is:
\be
\left(\alpha J \sqrt{\alpha^2 + \beta^2 (\check{K}^2-1)}\right)^{-1}\,.
\label{J}
\ee
Using the standard change of variables formula, it follows that for $J \in (0,F_0], \check{K} \geq 1$,
the conditional bivariate PDF of the pair $(\underline{F}^s_T, \check{F}^s_T)$ is given by:
\bea
\mathbb{Q} \{\un{F}^s_T \in dJ, \check{F}^s_T \in d\check{K}|N_T = 0, \un{F}^s_t=\un{F}, \check{F}_t = \check{F}\} = f(J,\check{K};w,T-t) dJ d\check{K} \nn\\
 f(J,\check{K};w,T-t) \equiv \sqrt{ \frac{2}{\pi (T - t)^3}} \frac{\left(\check{k}(\check{K}) - j(J) + w\right)e^{- \frac{\left(\check{k}(\check{K})- j(J) + w \right)^2}{2(T-t)}}}{\alpha J \sqrt{ \alpha^2 + \beta^2 (\check{K}^2-1)}}
\,,
\eea
and
\be
w =  j(\un{F}) + \check{k}(\check{F}).
\label{Wt}
\ee
Note that $w=Z_t$, and the reason we use $w$ is that it is written on market observables $\check{F}$ and $\ul{F}$ while $Z_t$ is not.

Let $F^s_T = \un{F}^s_T \check{F}^s_T$ be the forward price at $T$
conditional on survival to $T$.
The bivariate PDF of the pair $(\underline{F}^s_T, \check{F}^s_T)$
can be used calculate the conditional transition PDF of $F^s_T$:
\bea
 \mathbb{Q} \{F^s_T \in dF|N_T = 0, \un{F}^s_t=\un{F}, \check{F}_t = \check{F}\} = g(F;w,T-t) dF \,,
\eea
where
\bea
 g(F;w,T-t) & = & \int_0^{F_0}  f \left(J, \frac{F}{J};w,T-t \right) dJ
\label{Fpdf} \\
 & = & \int_0^{F_0} 
 \sqrt{ \frac{2}{\pi (T - t)^3}} \frac{\left(k \left(\frac{F}{J} \right) - j(J) + w\right)
e^{- \frac{\left(k \left( \frac{F}{J} \right)- j(J) + w \right)^2}{2(T-t)}}}{\alpha J \sqrt{ \alpha^2 + \beta^2 \left[\left( \frac{F}{J}\right)^2-1 \right]}}
 dJ,
\nn
\eea
and $w$ is given in (\ref{Wt}).
When $F$ is only conditioned on surviving to $t$ rather than to $T$,
the transition PDF's of both $(\ul{F}_T,\check{F}_T)$ and $F_T$ are just given by the product of 
their corresponding transition PDF conditioned on survival to $T$ and the probability of further surviving to $T$,
 which is  $e^{-\frac{\beta^2}{2} (T-t)}$:
\bea
\mathbb{Q} \{\un{F}_T \in dJ, \check{F}_T \in d\check{K}|N_t = 0, \un{F}^s_t=\un{F}, \check{F}_t = \check{F}\} &=& f(J,\check{K};w,T-t) e^{-\frac{\beta^2}{2} (T-t)} dJ d\check{K}\,,
\nn\\
  \mathbb{Q} \{F_T \in dF|N_t=0, \un{F}^s_t=\un{F}, \check{F}_t = \check{F}\}
  &=& g(F;w,T-t)e^{-\frac{\beta^2}{2} (T-t)}dF\,.
\eea
The PDF of $F_T$  is an integral over a bounded domain and it cannot be simplified further.
We will find that
when common payoffs are integrated against this PDF, additional quadratures are not introduced.
It is for this reason that we consider the process $F$ to be tractable.

There are two similar constructions of a non-negative martingale which also
use jump to default.  
The cumulative hazard process of $N$  is 
$\Lambda_t = e^{\frac{\beta^2}{2} t}$
which is deterministic.
Suppose instead that the  cumulative hazard process  is 
$\hat{\Lambda}_t = e^{-\alpha \un{Z}_t}$, which is random.
Let  $\hat{N}$ denote the corresponding  counting process and 
let $\hat{F}$ denote the desired non-negative martingale:
  \be
\hat{F}_t
= F_0
 e^{- \frac{\beta^2}{2}t}
e^{\beta \check{Z}_t}_{\beta - \alpha}
\m1_{N_t = 0}\,,
\qquad t \geq t_0
\label{new10f}
\ee
is a non-negative martingale started at $F_0 > 0$.
Since $\un{Z}_0 = 0$, this process start off with no chance of 
jumping to zero but soon endures the possibility of such a default. 
More generally, one can start the process $\un{Z}$ at some non-positive number $m_0 \leq 0$ and rename the process $\un{Z}$ to say $m$
since $Z_0$ is still zero.
Since $\check{Z}_t = Z_t - m_t$ starts at $-m_0 >0$,  
one must then also adjust its origin:
\be
F_t
= F_0 
 e^{- \frac{\beta^2}{2}t}
e^{\beta (\check{Z}_t + m_0)}_{\beta - \alpha}
\m1_{\hat{N}_t = 0},
\qquad t \geq t_0
\label{new11f}
\ee

There is yet another construction of a non-negative martingale possibly jumping to zero.  
Now suppose that the cumulative hazard process of $N$  is 
$\tilde{\Lambda}_t = e^{-\alpha \un{Z}_t + \frac{\beta^2}{2} t}$, where we return to 
 $\check{Z}_t = Z_t - \un{Z}_t$ with $\un{Z}_ 0 = 0$.
Let  $\tilde{N}$ denote the corresponding  counting process and 
let $\tilde{F}$ denote the desired non-negative martingale:
  \be
\tilde{F}_t
= F_0
e^{\beta \check{Z}_t}_{\beta - \alpha}
\m1_{\tilde{N}_t = 0},
\qquad t \geq t_0
\label{new12f}
\ee
is a non-negative martingale started at $F_0 > 0$.
This process is convenient if an event happens
at the first passage time $\tau$ of 
$F$ to a constant  upper barrier 
$H = e^{\beta h}_{\beta - \alpha}$ where $h > 0$.
In this case, $\tau$ is also the first passage time of 
$\check{Z}$ to $h$.
Since $\frac{F}{F_0}$ is a martingale started at one, the
bivariate  Laplace transform of $ \un{Z}_{\tau}$ and $\tau$ becomes known:
\be
= E e^{\alpha  \un{Z}_{\tau} - \frac{\beta^2}{2} \tau}
\frac{1}{e^{\beta h}_{\beta - \alpha}}.
\label{blt}
\ee

One can develop yet other tractable constructions of non-negative martingales 
by altering the cumulative hazard process yet again
and compensating by coordinate change as was done above.

\section{Application in Option Pricing}
In risk neutral measure, non-arbitrage insures that the expected payoff of a security is equal to its current price. In this section we show how our model can be applied in derivative pricing assuming the underlying asset follows the dynamics of either the sub-martingale Eqn (\ref{submgl}) or the martingale Eqn (\ref{new1f}) in risk neutral measure. The former is used for derivatives written on spot price of a security while the latter is for future price of a security. Since the two processes only differ by the inclusion of a jump to default process, the pricing formulas for them are quite close. For this reason, we only present the derivation of pricing for the martingale dynamics. The results for the sub-martingale dynamics are labelled by subscripts for clarification. Note since our model tracks the asset's running minimum and drawup rate, it is especially useful in pricing barrier type of path-dependent options.

\subsection{One-Touch with a lower barrier}\label{OT}
We first price a One-Touch with a lower barrier. A One-Touch option pays one dollar if the underlying asset's price touches the lower barrier price before maturity, and otherwise expires worthless. Assuming that the present time is $t$ and the underlying asset has not defaulted ($N_t = 0$). The price of a One-Touch with a lower barrier $L$ and maturity $T$ is
\bea
OT_t(L,T) &=& \m1 _{\underline{F}_t \le L} + \m1_{\underline{F}_t >L}\cdot \left(\m1_{N_T = 0} E_t\left[ \m1_{\underline{F}_T\le L} \right] +\m1_{N_T\ne 0}\right)\nn\\
&=& \m1 _{\underline{F}_t \le L} + \m1_{\underline{F}_t >L} \cdot \left(e^{-\frac{\beta^2(T-t)}{2}}E_t\left[ \m1_{\underline{Z}_T\le \frac{\ln L-\ln F_0}{\alpha}} \right] +1-e^{-\frac{\beta^2(T-t)}{2}} \right)\,,
\eea
to get the second line, $\underline F _T = F_0 e^{\alpha \underline{Z} _T}$ has been used. After substituting the transition PDF on $\underline{Z}_T$ one obtains
\bea
OT_t(L, T) = \m1_{\underline F_t \le L} + \m1_{\underline F_t >L} \cdot \left(e^{-\frac{\beta^2(T-t)}{2}} \left[2N\left(\frac{\frac{\ln L-\ln F_0}{\alpha} - w}{\sqrt{T-t}}\right) -1\right]+1\right)\,,
\eea
where $w$ is given in (\ref{Wt}) and $N$ is the standard normal distribution function. Taking $\alpha =1$ the price of the One-Touch reduces to that in \cite{c14}. This is because essentially the payoff of a One-Touch option is only determined by the minimum of the underlying, which is driven by the running minimum of a Brownian motion in both cases. 

A One-Touch written on spot price can be priced similarly with Eqn (\ref{submgl}), which is equivalent to dropping the probability induced by the jump to default process in Eqn (\ref{new1f}). The price is then given by
\bea
OT^{\rm{Spot}}_t(L, T) &=&\m1 _{\underline{F}_t \le L} + \m1_{\underline{F}_t >L}\cdot E_t\left[ \m1_{\underline{F}_T\le L} \right] \nn\\
&=& \m1_{\underline F_t \le L} + \m1_{\underline F_t >L} \cdot 2N\left(\frac{\frac{\ln L-\ln F_0}{\alpha} - w}{\sqrt{T-t}}\right) \,.
\eea

\subsection{Lookback option}
A lookback call option matures at $T$ with a floating strike price pays off the difference between the terminal value of the asset and its minimum, namely the terminal drawup. If default happens $(N_T\ne 0)$, the option expires worthless $(F_T=\ul{F}_T)$. So under the martingale (\ref{new1f}) the value of this option at maturity is then 
\bea
LC_{float,t}&=& \m1_{N_T=0} E_t\left[ F_T -\underline{F}_T \right] = \m1_{N_T=0} E_t\left[ \ul{F}_T\left(\check{F}_T-1\right)\right] \nn\\
&=& \m1_{N_T=0} E_t\left[ F_0e^{\alpha \ul{Z}_T}\left(e^{\beta \check{Z}_T}_{\beta-\alpha}-1\right)\right]\,.\label{lc1}
\eea
The expectation value in Eqn (\ref{lc1}) can be evaluated using the bivariate transition PDF of $(\ul{Z}_T,\check{Z}_T)$  the in Eqn (\ref{jden}) if the security runs into a new minimum after $t$, or otherwise the univariate transition PDF of $\check{Z}_T$ in Eqn(\ref{jden1}) if $\ul{Z}_T=\ul{Z}_t$:
\bea
E_t\left[ F_0 e^{\alpha \ul{Z}_T}\left(e^{\beta \check{Z}_T}_{\beta-\alpha}-1\right)\right]
 &=& F_0 \int^{\ul{Z}_t}_{-\infty}dj\int_0^{\infty}d\check{k} \sqrt{ \frac{2}{\pi (T - t)^3}} (\check{k}-j + w)e^{- \frac{(\check{k}-j+w)^2}{2(T-t)}} e^{\alpha j}\left(e^{\beta \check{k}}_{\beta-\alpha}-1\right)\nn\\
&+&F_0\int_0^{\infty} d\check{k} \sqrt{ \frac{2}{\pi (T - t)}} \left( e^{-\frac{\check{k}^2}{2(T-t)}} - e^{-\frac{(\check{k}+\check{w})^2}{2(T-t)}} \right)  e^{\alpha \ul{w}}\left(e^{\beta \check{k}}_{\beta-\alpha}-1\right)\,,
\eea
where $\check{w}=w-\ul{w}$. By working out the integral we obtain the price of this option evaluated at $t$ 
\bea
LC_{float, t}&=&F_0 e^{\alpha \ul{w}}\bigg[ \frac{\alpha}{\beta}e^{\beta \check{w}}N\left(\frac{-\check{w}-\beta(T-t)}{\sqrt{T-t}}\right) -\frac{\alpha}{\beta}e^{-\beta \check{w}}N\left(\frac{-\check{w}+\beta(T-t)}{\sqrt{T-t}}\right) \nn\\
&+& \frac{\beta +\alpha}{\beta}N\left(\beta \sqrt{T-t}\right)+\frac{\beta -\alpha}{\beta}N\left(-\beta \sqrt{T-t}\right) +e^{-\frac{\beta^2 (T-t)}2} \left(2N\left( \frac{-\check{w}}{\sqrt{T-t}}\right)  -1\right)
\nn\\
&-&2e^{\alpha \check{w}+\frac{(\alpha^2-\beta^2)(T-t)}2}N\left(\frac{-\check{w}-\alpha(T-t)}{\sqrt{T-t}}\right)\bigg]\,.
\label{lc}
\eea
A lookback call option on spot price can be priced the same way:
\bea
LC^{\rm{Spot}}_{float,t} &=&  E_t\left[ F_T -\underline{F}_T \right] = E_t\left[ F_0e^{\alpha \ul{Z}_T}\left(e^{\beta \check{Z}_T}_{\beta-\alpha}-1\right)\right]\nn\\
&=& e^{\frac{\beta^2 (T-t)}2}LC_{float,t}\,.\label{fxlc}
\eea
If we instead consider a lookback option with a fixed strike price, then the payoff is determined by the minimum/maximum for a put/call lookback option at maturity. Since Eqn (\ref{new1f}) tracks minimum and drawup, it can also be used to evaluate a lookback put option with fixed price. The price is given by
\bea
LP_{fixed, t}(K,T)&=&\m1_{N_T=0}E_t\left[(K-\ul{F}_T)^+\right]+\m1_{N_T\ne 0}\cdot K\,,\nn\\
LP^{\rm{Spot}}_{fixed, t}(K,T)&=&E_t\left[(K-\ul{F}_T)^+\right]\,,
\eea
where $K$ is the strike price. This can be evaluated by integrating the price of a one-touch barrier with respect to the barrier, so we will not carry out the derivation for simplicity.

We can also engineer another derivative analogous to a lookback call option with a floating strike price, which pays off the ratio between the terminal price and the minimum price before maturity. Since the underlying asset can default ($F_T=\ul{F}_T=0$), we assume the payoff is zero in that case. The price of this option is given by
\bea
LC^*_{float,t}&=&\m1_{N_T=0} E_t\left[\frac{F_T-\ul{F}_T}{\ul{F}_T}\right]=\m1_{N_T=0} \left( E_t\left[ \frac{F_T}{\ul{F}_T}\right] -1 \right)
\nn\\
&=&\m1_{N_T=0} \left( E_t\left[ e^{\beta \check{Z}_T}_{\beta-\alpha}\right] -1 \right)\,,
\eea
and the expectation can be evaluated with the bivariate PDF:
\bea
E_t\left[ e^{\beta \check{Z}_T}_{\beta-\alpha}\right]&=&
\int^{\ul{Z}_t}_{-\infty}dj\int_0^{\infty}d\check{k} \sqrt{ \frac{2}{\pi (T - t)^3}} (\check{k}-j + w)e^{- \frac{(\check{k}-j+w)^2}{2(T-t)}} e^{\beta \check{k}}_{\beta-\alpha} \nn\\
&+& \int_0^{\infty} d\check{k} \sqrt{ \frac{2}{\pi (T - t)}} \left( e^{-\frac{\check{k}^2}{2(T-t)}} - e^{-\frac{(\check{k}+\check{w})^2}{2(T-t)}} \right)  \left(e^{\beta \check{k}}_{\beta-\alpha}-1\right)
\eea
which can be evaluated similar to Eqn (\ref{lc}),
\bea
LC_{float,t}^*&=&\frac{\beta +\alpha}{\beta}N\left(\beta \sqrt{T-t}\right)+\frac{\beta -\alpha}{\beta}N\left(-\beta \sqrt{T-t}\right)-1\,.
\eea
Note that the value of $LC_{float,t}^*$ is unitless, since the option is written on the drawup ratio. If there is a size associated to the underlying security, it can be multiplied to $LC_{float,t}^*$ which gives it a dollar amount. As in Eqn (\ref{fxlc}), the price for this derivative on spot price is 
\bea
LC^{*\,\rm{Spot}}_{float,t}=e^{\frac{\beta^2 (T-t)}2}LC^*_{float,t}\,.
\eea

\subsection{Vanilla and Down-and-In Call}
Now we price a Down-and-In Call (DIC) option which becomes from worthless to a vanilla call if the lower barrier is hit before maturity. A vanilla call can be viewed as a special case of a Down-and-In barrier call (DIC) with the lower barrier has been hit prior to presence. The value of a DIC option written on $F_t$ is given by 
\be
{\rm{DIC}}_t(L, K, T) =\m1_{\un{F}_t \le L}\cdot \m1_{N_T = 0}\cdot C_t(K,T)+ \m1_{\un{F}_t > L}\cdot \m1_{N_T = 0}\cdot E_t \left[ \m1_{\un{F}_T \le L} (F_T - K)^+\right]\,,\label{dic}
\ee
where $L$ is the barrier, $K$ is the strike price, $T$ is maturity and $C_t$ is a vanilla call priced at $t$. Note setting $L ={F}_0$ reduces the DIC to a vanilla call. As implied by Eqn (\ref{dic}) if default happens ($N_T\ne 0$), the option becomes worthless. To evaluate the expectation value of the second term in (\ref{dic}), we once again apply the bivariate transition PDF:
\bea
E_t \left[\m1_{\un{F}_T \le L} (F_T - K)^+\right] &=& E_t\left[\m1_{\underline{Z}_T \le \frac 1{\alpha}\ln{\frac L{F_0}}} (F_0 e^{\alpha\underline{Z}_T}\ew - K)^+\right]\nn\\
&=& \int_{-\infty}^{\frac 1{\alpha}\ln{\frac L{F_0}}} dj \int_{k^*}^{\infty}d\check{k} \sqrt{ \frac{2}{\pi (T - t)^3}} (\check{k}-j + w)e^{- \frac{(\check{k}-j+w)^2}{2(T-t)}} \left(F_0 e^{\alpha j}e^{\beta \check{k}}_{\beta-\alpha}-K\right)\,,
\eea
where $k^*(j)$ is determined by
\bea
k^*={\rm{max}}\left(f^{-1}\left(\frac S {F_0 e^{\alpha j}}\right),0\right)\,, \qquad f(x)=e^{\beta x}_{\beta-\alpha}\,.
\eea
For the dependence of $k^*$ on $j$, the integral above cannot be obtained in closed form, a similar situation as in \cite{c14}. Nonetheless, the result can be further simplified as 
\bea
E_t \left[\m1_{\un{F}_T \le L} (F_T - K)^+\right]
&=& F_0  \int_{-\infty}^{\frac 1{\alpha}\ln{\frac L{F_0}}} dj e^{\alpha j+\frac{\beta^2(T-t)}2}\bigg[(\beta+\alpha)e^{\beta(j-w)}N\left(\frac{j-w-k^*+\beta (T-t)}{\sqrt{T-t}}\right) \nn\\
&&\qquad -(\beta-\alpha)e^{-\beta(j-w)}N\left(\frac{j-w-k^*-\beta (T-t)}{\sqrt{T-t}}\right) \bigg]\,, 
\eea
which gives rise to the value of the DIC option 
After replacing $Z_t$ with the market observable $w$, we now have the price for the DIC option:
\bea
{\rm{DIC_t}}(L, K, T)
		&=&\m1_{\underline{F}_t \le L}C_t(K,T) +\m1_{\un{F}_t > L}  F_0  \int_{-\infty}^{\frac 1{\alpha}\ln{\frac L{F_0}}} dj e^{\alpha j}\bigg[(\beta+\alpha)e^{\beta(j-w)}N\left(\frac{j-w-k^*+\beta (T-t)}{\sqrt{T-t}}\right) \nn\\
&&\qquad \qquad -(\beta-\alpha)e^{-\beta(j-w)}N\left(\frac{j-w-k^*-\beta (T-t)}{\sqrt{T-t}}\right) \bigg]\,.
\label{dic2}
	\eea
In the special case when $L = F_0$, the DIC option reduces to a vanilla call with a price of
       \bea
		C_t( K, T) &=&  F_0  \int_{-\infty}^{0} dj e^{\alpha j}\bigg[(\beta+\alpha)e^{\beta(j-w)}N\left(\frac{j-w-k^*+\beta (T-t)}{\sqrt{T-t}}\right)\nn\\
&&\quad\quad  -(\beta-\alpha)e^{-\beta(j-w)}N\left(\frac{j-w-k^*-\beta (T-t)}{\sqrt{T-t}}\right) \bigg]\,, 
\label{call}
	\eea
which completes the pricing of a DIC option on Eqn (\ref{new1f}). For a DIC option on spot price, Eqn (\ref{dic}) becomes
\be
{\rm{DIC}}^{\rm{Spot}}_t(L, K, T) =\m1_{\un{F}_t \le L}\cdot C^{\rm{FX\,\, Spot}}_t(K,T)+ \m1_{\un{F}_t > L}\cdot E_t \left[ \m1_{\un{F}_T \le L} (F_T - K)^+\right]\,,
\ee
which leads to slight modification on both Eqn (\ref{dic2}) and Eqn (\ref{call}), and the results are
\bea
{\rm{DIC}}^{\rm{Spot}}_t(L, K, T)
&=&\m1_{\underline{F}_t \le L}C^{\rm{Spot}}_t(K,T) \nn\\
&+&\m1_{\un{F}_t > L}  F_0 e^{\frac{\beta^2 (T-t)}2}  \int_{-\infty}^{\frac 1{\alpha}\ln{\frac L{F_0}}} dj e^{\alpha j}\bigg[(\beta+\alpha)e^{\beta(j-w)}N\left(\frac{j-w-k^*+\beta (T-t)}{\sqrt{T-t}}\right) \nn\\
&&\qquad \qquad -(\beta-\alpha)e^{-\beta(j-w)}N\left(\frac{j-w-k^*-\beta (T-t)}{\sqrt{T-t}}\right) \bigg]\,,\nn\\
C^{\rm{ Spot}}_t( K, T) &=&  F_0 e^{\frac{\beta^2 (T-t)}2}  \int_{-\infty}^{0} dj e^{\alpha j}\bigg[(\beta+\alpha)e^{\beta(j-w)}N\left(\frac{j-w-k^*+\beta (T-t)}{\sqrt{T-t}}\right)\nn\\
&&\quad\quad  -(\beta-\alpha)e^{-\beta(j-w)}N\left(\frac{j-w-k^*-\beta (T-t)}{\sqrt{T-t}}\right) \bigg]\,. 
\eea

Before closing this section, we would like to point out that Eqn (\ref{dic2}) is related to several options. For instance, when $\alpha=1$ and $\beta = 0$ the result reduces to that in \cite{c14}. In the special case of a zero strike DIC option ($K=0$), Eqn (\ref{dic2}) has closed form expressions:
\bea
&&DIC_t(L,0,T)=F_0\bigg[\left(\frac L{F_0}\right)^{\alpha+\beta}e^{-\beta w}N\left(\frac{\frac 1{\alpha}\ln \frac L{F_0}-w+\beta(T-t)}{\sqrt{T-t}}\right)\nn\\
 &&\quad\quad -2e^{\alpha w +\frac{(\alpha^2-\beta^2)(T-t)}2}N\left( \frac{\frac 1{\alpha}\ln \frac L{F_0}-w-\alpha(T-t)}{\sqrt{T-t}}\right)+\left(\frac L{F_0}\right)^{\alpha-\beta}e^{\beta w}N\left(\frac{\frac 1{\alpha}\ln \frac L{F_0}-w-\beta(T-t)}{\sqrt{T-t}}\right)\bigg].
\eea

\section{Summary and Extensions}

We proposed a three parameter continuous martingale
with state space $[0,\infty)$. This is done by first generating a process with a positive drift driven by the running minimum and drawup of a Brownian motion in the Az{\' e}ma-Yor setting, and adding a jump to default process. 
The process generalizes driftless Geometric  Brownian motion by adding two more parameters
while preserving its tractability. In particular, its running minimum and drawup rate (the ratio between level and running minimum) are both analytically tractable.
The three model parameters
$\alpha, \gamma$, and $\beta$
can be respectively interpreted as the instantaneous volatility
of the underlying at each new low,
at the initial time, and at infinitely high prices of the underlying.
The parameter $\alpha$ controls the implied volatility at
low strikes, while  the parameter $\beta$ controls the implied volatility at
high strikes.
So long as implied volatility is monotonic in strike price,
the  parameter $\gamma$ can be used to meet an at-the-money implied
volatility. It is shown that in certain limits, this new process can reduce to Geometric Brownian motion and the positive martingale given in \cite{c14}. We also presented the bivariate transition PDF of the process' running minimum and drawup rate. By utilizing the transition PDF, we priced several options assuming the dynamics are driven by the three parameter martingale in risk neutral measure. The options include a one-touch option with a lower barrier, lookback options with floating and fixed strike prices, vanilla call and a down-and-in call option.  

Since not all implied volatility slices are monotonic,
future research should be directed towards
extending the model by introducing either stochastic volatility or jumps.
One can also use the process without jump to default to model dynamics that involve a positive drift, for instance, the cumulative return of an investment strategy. 
Moreover, Girsanov's theorem can be used to remove the drift of $G$,
at which point a reflection principle becomes available.
In the interests of brevity, these extensions are best left for future research.

\section*{Acknowledgement}
We are grateful to Matthew Lorig, Vasily Strela,  Jane Yu, and especially 
 Travis Fisher,
 for their comments.
They are  not responsible for any errors.

\section*{Appendix}
\subsection*{1. More about $e^{\beta x}_{\beta - \alpha}$}
This technical appendix proves the result (\ref{swap}).
For $x \geq 0$,  $\alpha \geq 0$, and $\beta>0$,
our two parameter exponential function
is defined as:
 \be
e^{\beta x}_{\alpha - \beta} \equiv
\frac{\alpha + \beta}{2 \beta} e^{\beta x}
+
\frac{\alpha - \beta}{2 \beta} e^{-\beta x}.
\label{agex2}
\ee
Squaring this result implies that:
\be
\left( e^{ \beta x}_{\alpha - \beta} \right)^2 =
\left( \frac{\alpha + \beta}{2 \beta} \right)^2 e^{2\beta x}
+
\frac{\alpha^2 - \beta^2}{2 \beta^2}
+ \left( \frac{\alpha - \beta}{2 \beta} \right)^2e^{-2\beta x}.
\label{agex3}
\ee
Consider the cohort of (\ref{agex2}):
\be
e^{\beta x}_{\beta - \alpha} \equiv
\frac{\alpha + \beta}{2 \beta} e^{\beta x}
+
\frac{\beta - \alpha}{2 \beta} e^{-\beta x}.
\label{agex}
\ee
Squaring this cohort implies that:
\be
\left( e^{ \beta x}_{\beta - \alpha} \right)^2 =
\left( \frac{\alpha + \beta}{2 \beta} \right)^2 e^{2\beta x}
-
\frac{\alpha^2 - \beta^2}{2 \beta^2}
+ \left( \frac{\alpha - \beta}{2 \beta} \right)^2e^{-2\beta x}.
\label{agex1}
\ee
Subtracting (\ref{agex1}) from (\ref{agex3}) implies that:
\be
\left( e^{ \beta x}_{\alpha - \beta} \right)^2-
\left( e^{ \beta x}_{\beta - \alpha} \right)^2
=
\frac{\alpha^2 - \beta^2}{\beta^2}.
\label{diff}
\ee
Taking the positive square root of each side leads to the desired result:
\be
e^{\beta x}_{\alpha - \beta} =
\sqrt{ \left( e^{\beta x}_{\beta - \alpha} \right)^2 + \frac{\alpha^2 - \beta^2}{\beta^2}}.
\label{aswap}
\ee

\ed